\documentclass[10pt,letterpaper]{article}
\usepackage{opex3}
\usepackage[fleqn]{amsmath}
\usepackage{cite}

\begin{document}

\title{Multimode nanobeam cavities for nonlinear optics: high quality resonances separated by an octave}

\author{Sonia Buckley$^{1,*}$, Marina Radulaski$^{1}$, Jingyuan Linda Zhang$^{1}$, Jan Petykiewicz$^{1}$, Klaus Biermann$^{2}$, Jelena Vu\v{c}kovi\'{c}$^{1}$}

\address{$^1$Spilker Center for Engineering and Applied Sciences, Stanford University, Stanford CA 94305\\
$^2$Paul-Drude-Institut f\"{u}r Festk\"{o}rperelektronik, Hausvogteiplatz 5-7 D-10117, Berlin, Germany}

\email{*bucklesm@stanford.edu} 



\begin{abstract*}
We demonstrate the design, fabrication and characterization of nanobeam cavities with multiple higher order modes.  Designs with two high Q modes with frequency separations of an octave are introduced, and we fabricate such cavities exhibiting resonances with wavelength separations of up to 740 nm. \\
\\
\end{abstract*}


\section{Introduction}

On chip nanophotonic frequency conversion has made great progress within the past decade. In particular, $\chi^{(3)}$ waveguides and microcavities for frequency comb generation in the visible, telecom and out to the mid-IR \cite{delhaye_optical_2007,levy_cmos-compatible_2010,razzari_cmos-compatible_2010, witzens_silicon_2010, Schliesser2012}, Raman lasing \cite{takahashi_micrometre-scale_2013} and entangled photon pair generation
\cite{clemmen_continuous_2009} have all been demonstrated in a variety of materials. While there has also been similar work on $\chi^{(2)}$ nanophotonics, the performance of such on-chip devices still lags behind its $\chi^{(3)}$ counterparts, despite the relatively higher nonlinearity at lower powers. This discrepancy can in part be attributed to the difficulty in achieving the dispersion engineering necessary for meeting the phase-matching condition for photonic modes that are very far apart in frequency.  For microcavities with strong mode localization, this requires solving the challenging problem of engineering high quality factor (Q) modes with large frequency separations that maintain good spatial overlap \cite{rivoire_multiply_2011, rivoire_multiply_2011-1, thon_polychromatic_2010, zhang_ultra-high-q_2009}.  Realizing such low power, high efficiency devices could have applications in quantum information processing for on-chip frequency down-conversion of flying qubits emitted by solid state emitters such as InAs quantum dots (QDs) to telecommunications wavelengths \cite{mccutcheon_broadband_2009}, or up-conversion to the optimal frequency window for single photon detection \cite{langrock_highly_2005}. Additional applications include on chip low power spectroscopic sources, up-conversion of mid-IR to visible for imaging applications \cite{zhou_ultrasensitive_2013}, and for fundamental science studies such as strong coupling of single photons \cite{irvine_strong_2006} or single photon blockade \cite{majumdar_single-photon_2013}.

Photonic crystal cavities (PCCs) offer great promise for nonlinear frequency conversion, due to the high quality factors and low mode volumes.  Recently, Raman lasing in high Q silicon PCCs was demonstrated \cite{takahashi_micrometre-scale_2013}, while the $\chi^{(2)}$ processes of second harmonic generation (SHG) and sum frequency generation (SFG) have also been demonstrated in such cavities in III-V semiconductors \cite{mccutcheon_experimental_2007,rivoire_second_2009, rivoire_sum-frequency_2010, buckley_second_2013, Ota2013:Nanocavity-based, Ota2013:Self-frequency, Buckley2014a}, as well as in other materials such as lithium niobate \cite{Diziain2013}, SiC \cite{Yamada2014} and Si \cite{Galli2010}.  However, as described above, the difficulty in engineering cavities with modes that are far apart in frequency has limited the ability to increase the efficiency at low power levels.

1D planar nanobeam PCCs \cite{Notomi2008, Foresi1997, Lalanne2004, Sauvan2005, Deotare2009} in particular show promise for nonlinear frequency conversion.  Q factors of $>10^9$ have been simulated in these structures \cite{Quan2011}, with Q factors as high as $7.5\times10^5$ demonstrated on a Si platform \cite{Deotare2009}.  Another advantage of 1D periodic structures is that they maintain a photonic band gap with lower index contrast than is possible for 2D planar photonic crystals, and therefore can still exhibit high Q factor modes on very low refractive index materials such as SiO$_2$ \cite{Gong2010}, diamond \cite{Babinec2011}, Si$_3$N$_4$ \cite{Khan2011}, sputter coated AlN \cite{Pernice2012} and SiC \cite{Radulaski2013}. Moreover, in general a larger photonic band gap can be achieved in 1D than in 2D periodic structures, as one is only optimizing the band gap in one dimension.  Nanobeams for four wave mixing \cite{lin_high-efficiency_2013},  terahertz frequency generation \cite{burgess_design_2009} and broadband frequency conversion of single photons \cite{mccutcheon_broadband_2009} have been proposed.  Four wave mixing in coupled nanobeam cavities has been demonstrated \cite{azzini_stimulated_2013}, while the TE/TM nanobeam cavities proposed for broadband frequency conversion of single photons have been fabricated in silicon \cite{zhang_ultra-high-q_2009, mccutcheon_high-q_2011}, but have yet to be implemented in high $\chi^{(2)}$ materials.  Crossed nanobeam PCCs \cite{rivoire_multiply_2011, rivoire_multiply_2011-1} offer another method for increasing the frequency separation of the resonances, as the ability to individually select their wavelengths is significantly improved and the thickness of the wafer is less critical; however such cavities work well only when the frequency separation of the nanobeams is not too large.

Here we demonstrate that the combination of lower and higher order modes in single nanobeam PCCs can be used for nonlinear frequency conversion.  We first design and optimize nanobeam cavities to obtain the highest Q factor modes. Next, we describe the rubric for optimizing Q versus frequency separation. We describe our design for nanobeam cavities with resonances separated by an octave, and evaluate this and several other modes in terms of their potential for frequency conversion.  Finally, we characterize optimized nanobeam cavities fabricated in (111)B oriented GaAs membranes via fiber taper and cross polarized reflectivity. We note that we have recently demonstrated sum frequency generation in such optimized structures \cite{Buckley2014:Nonlinear}.

\section{Optimization of higher order modes of the nanobeam}
\label{section:optimization}

\begin{figure}
\includegraphics[width = 10cm]{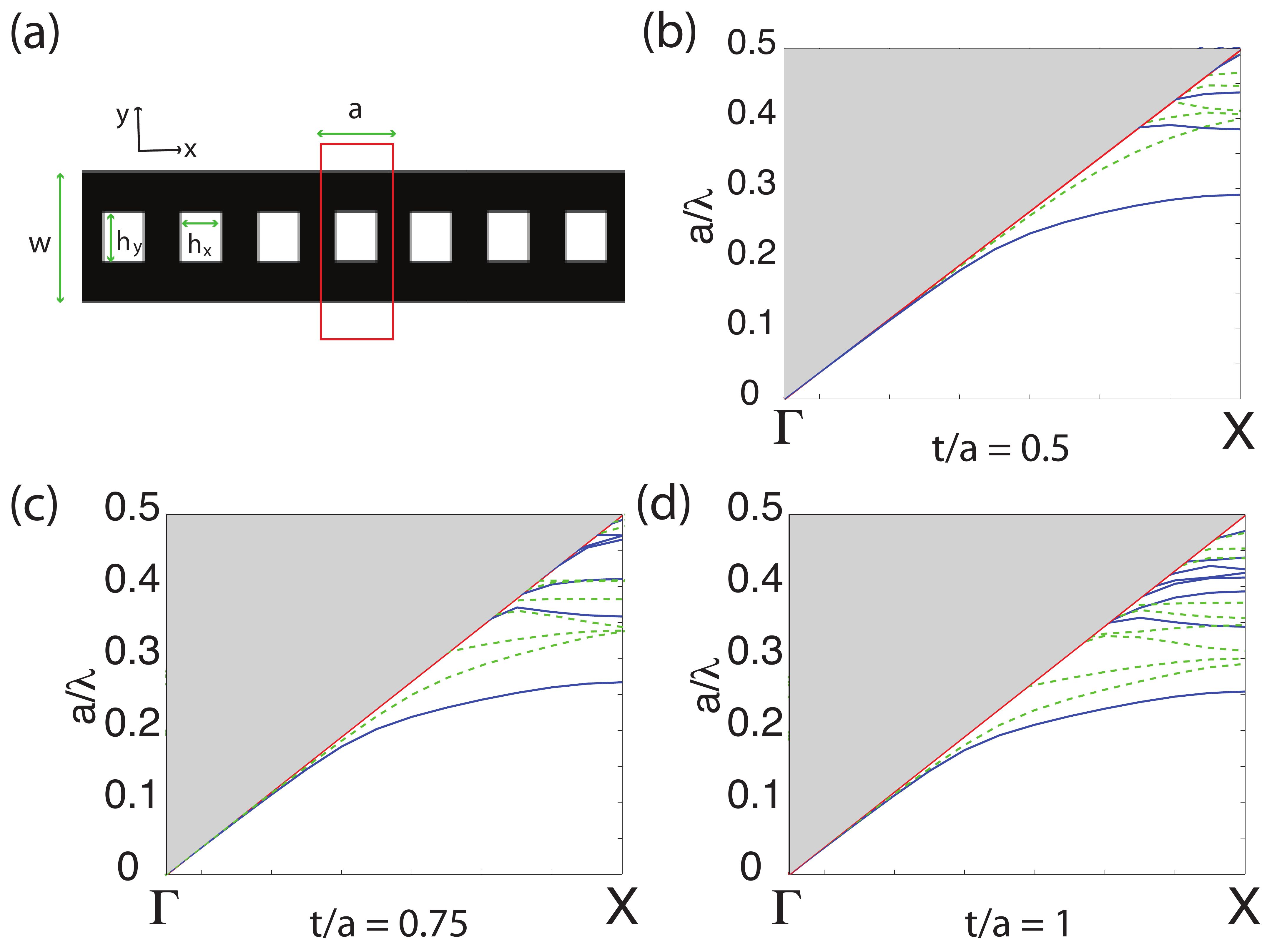}
\centering
{\caption{(a) Schematic of a nanobeam photonic crystal with parameters and coordinate directions indicated. (b)-(d) Band diagram for nanobeam with $h_x = 0.7 w$, $h_y= 0.5 a$, n = 3.34 for membrane thickness $t/a$ = 0.5, $t/a$ = 0.75 and $t/a$ = 1, respectively.  The blue solid lines show TE modes, the green dashed lines show TM modes.}
\label{fig:band_diags}}
\end{figure}

\begin{figure}
\includegraphics[width = 11cm]{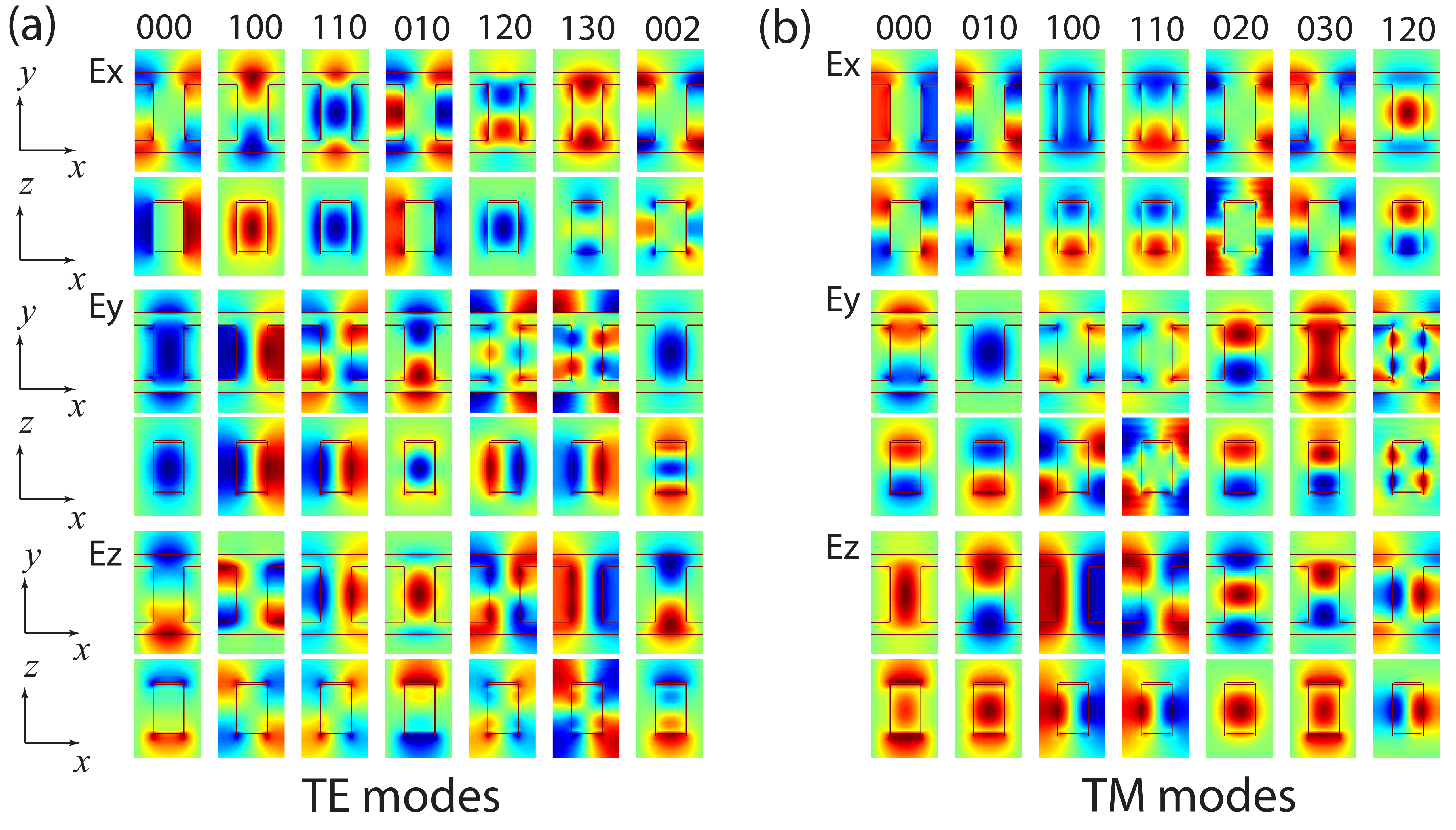}
\centering
{\caption{Field profiles for the first seven modes of the nanobeam with thickness $t/a$ = 0.75
shown in Fig \ref{fig:band_diags} (c).  The coordinate axes indicated in part (a) are repeated for part (b).}
\label{fig:band_diag_modes2}}
\end{figure}

We first consider a nanobeam with lattice constant $a$, beam width $w =1.2a$, hole width $h_x = 0.7w$ and hole height $h_y = 0.5w$ as shown in Fig. \ref{fig:band_diags} (a). The bands of this structure for different membrane thicknesses $t/a$, simulated with MIT photonic bands (MPB) \cite{johnson_block-iterative_2001}, are shown for refractive index of 3.34 (GaAs at 1.8 $\mu$m) in Fig. \ref{fig:band_diags} (b)-(d).  Transverse electric like (TE-like, i.e., vertically even) modes are indicated by solid blue lines and transverse magnetic like (TM-like, i.e., vertically odd) modes by green dashed lines, and the light line indicated by the red line.  These polarizations occur due to mirror symmetry in the central $xy$ plane \cite{Joannopoulos2011}, where $z$ is the surface normal direction.  As we will discuss later, a gentle perturbation in the periodic structure can localize one of these bands to the perturbed region; therefore studying the unperturbed band structure gives us information on the potential cavity modes of the system.  Our band diagram simulations did not include the change in refractive index with frequency. Including this dispersion effect will slightly shift the higher order bands to lower frequencies.  Dispersion effects were included in the later cavity simulations. As can be seen from Fig. \ref{fig:band_diags} (b)-(d), as the thickness of the membrane is increased, the bands move to lower frequencies beneath the light line. Increasing the refractive index or the width of the nanobeam will induce a similar trend, although the slope of the bands and their relative positions may also be changed.  As bands above the light line are all leaky, we are most interested in those close to or below the light line at the X-point. For $t/a$ = 0.5, the TE bands are well spaced, and can be localized by a gentle adiabatic taper to form high Q cavity modes. However, the maximum separation between high Q modes that can be attained is limited by the frequency separation between the lowest band beneath the light line and the frequency of the light line at the $X$ point. As the thickness is increased to $t/a$ = 0.75, the bands are all pushed lower beneath the light line, increasing the potential separation between modes, but are additionally pushed closer together. This increase in the mode density will lead to lower Q cavity modes, due to coupling between modes. This is even more obvious in the case of $t/a = 1$.  Fig. \ref{fig:band_diag_modes2} shows the field distribution within a single unit cell for the first seven bands at the $k = X$ point, for the $t/a = 0.75$ band diagram and for both TE and TM modes.  TE$_{ijk}$ (TM$_{ijk}$) bands are named according to the number of nodes $i$, $j$ and $k$ in the $x$, $y$ and $z$ directions for the $H_z$ ($E_z$) field component.

\begin{figure}
\includegraphics[width = 14cm]{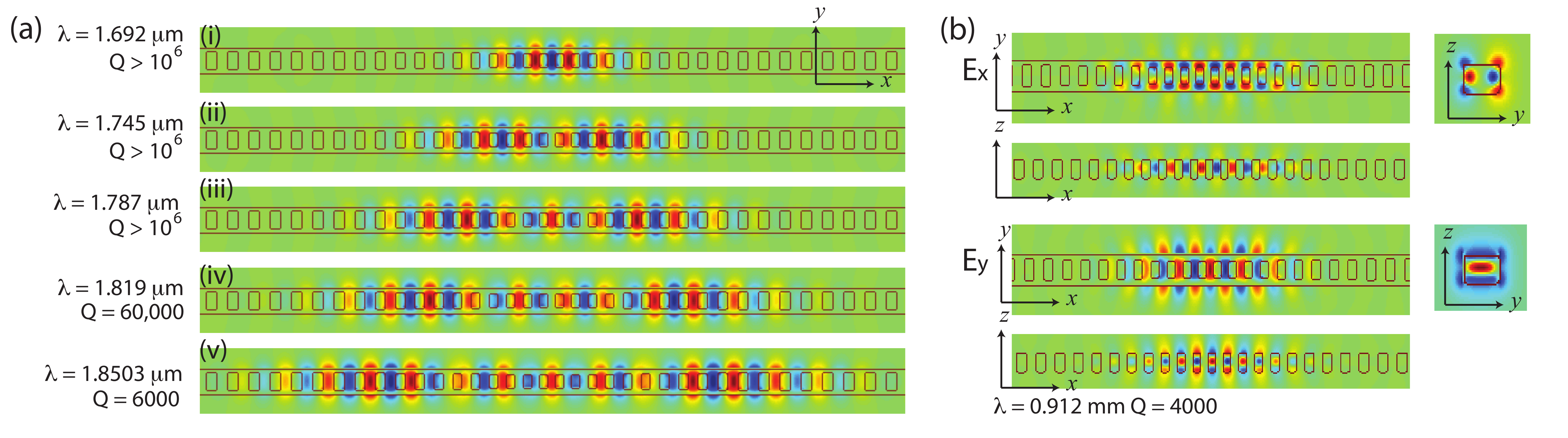}
\centering
{\caption{(a) The $E_y$ field component of the first to fifth order TE modes localized by the presence of the cavity from the TE$_{000}$ band in order of increasing frequency for $t/a$ = 0.75. (b) The $E_x$ and $E_y$ field components of the first localized TE$_{002}$ mode.}
\label{fig:SHG_mode1}}
\end{figure}

We next show how we can confine lower and higher order cavity modes with high Q (we will show how these can be used for nonlinear frequency conversion in Section \ref{section:frequency}).  Cavity simulations were performed using a home-built finite-difference time domain (FDTD) solver run on graphical processing units (GPUs).  First, we choose to confine modes from the TE$_{000}$ and TE$_{002}$ bands, as these have an appropriate frequency separation for the case of SHG and, as we will show in Section \ref{section:frequency} and Appendix \ref{app:100_vs_111}, the symmetry of the modes does not prohibit nonlinear interaction in III-V semiconductors in the (110) and (111) crystal orientations.  The confinement is done using a linear tapering of $h_x$, $h_y$ and $a$, with each reduced to a minimum of 0.8 of the value in the untapered region.  The first five modes confined from the TE$_{000}$ band for a membrane thickness of $t/a = 0.75$, beam width $w/a = 1.2$, $h_x = 0.7 w$ and $h_y = 0.5 a$ are shown in Fig. \ref{fig:SHG_mode1} (a). The modes alternate even and odd with respect to $x = 0$ and have decreasing Q factor and increasing extent (mode volume) with decreasing frequency. The mode volumes are 0.6$(\frac{\lambda}{n})^3$, 1.1$(\frac{\lambda}{n})^3$, 1.3$(\frac{\lambda}{n})^3$, 1.6$(\frac{\lambda}{n})^3$ and 1.9$(\frac{\lambda}{n})^3$.  Similarly for the TE$_{002}$ band, several modes can be localized, we obtain a Q factor of 4000 and a mode volume of 4$(\frac{\lambda_2}{n})^3$ for the highest frequency mode.   The refractive index for this mode at 900 nm is 3.5.  

\begin{figure}
\includegraphics[width = 14cm]{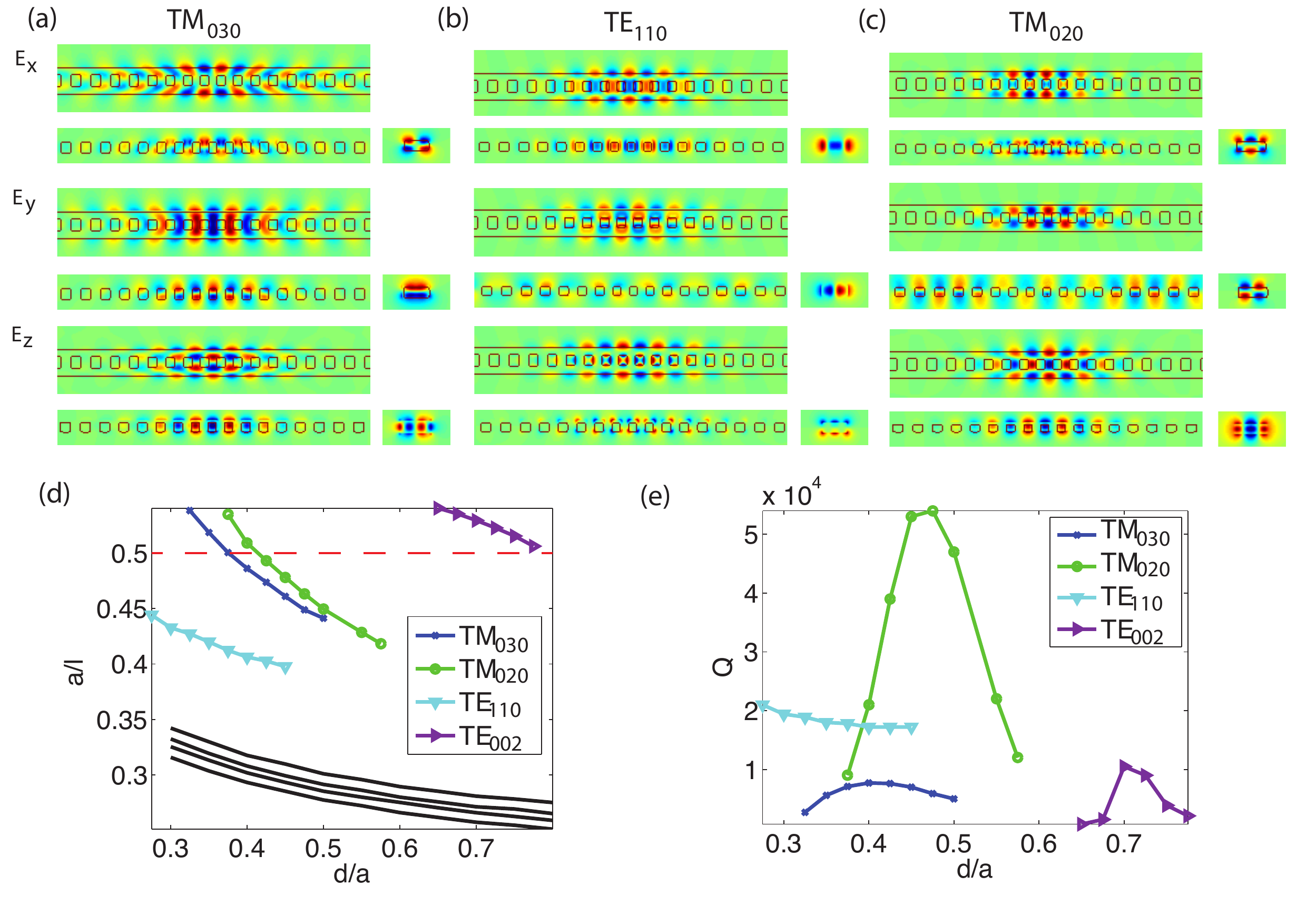}
\centering
{\caption{The $E_x$, $E_y$ and $E_z$ field profiles of the localized
(a) TM$_{030}$, (b) TE$_{110}$ mode and (c) TM$_{020}$ mode. (d) The mode frequency versus
membrane thickness $t/a$ for the nanobeam with $h_x=0.5w$, $h_y=0.5a$.  The solid
black lines indicate the first five localized modes of the TE$_{000}$ mode. The red
dashed line indicates the position of the light line at the X point.  (e) Q factor
versus membrane thickness.}
\label{fig:DFG_modes}}
\end{figure}

Many other modes can be localized for use in nonlinear frequency conversion.  Fig. \ref{fig:DFG_modes} shows three such modes for $h_x = 0.5w$ and $h_y = 0.5a$.  The three modes shown are the (a) TM$_{030}$, (b) TE$_{110}$ and (c) TM$_{020}$.  The frequency and Q factor versus membrane thickness $t/a$ is shown in part (c) and (d) for these modes, as well as for the TE$_{002}$ mode shown in the previous figure.  The maximum Q factor for each mode occurs at a different value of $t/a$, as shown in Fig. \ref{fig:DFG_modes} (e). The Q of the mode confined from the TE$_{002}$ band can be increased to $\approx 10,000$.  The mode is confined with high Q until the frequency approaches the light line at the X point (shown in the the red dashed line in Fig. \ref{fig:DFG_modes} (d). The Q decreases soon after this for all modes. 

\section{Frequency conversion in nanobeam cavities}
\label{section:frequency}

To facilitate high efficiency frequency conversion, it is desirable to choose the modes participating in the nonlinear frequency conversion process to have (1) well defined input/output modes; (2) high nonlinear overlap; and (3) high Q factors.  For a three wave mixing process such as SHG, SFG or difference frequency generation (DFG), defining three modes with high Q and high nonlinear overlap will lead to the highest conversion efficiency (this is reduced to 2 modes for the SHG where 2 of the 3 waves are degenerate). The conversion efficiency at low powers will be proportional to the nonlinear overlap between the three modes, given by \cite{Rodriguez2007, Burgess2009:Difference-frequency,burgess_design_2009}
\begin{align}
\label{eq:beta}
\left|\beta\right|^2 = \left|\frac{1}{4} \frac{\iiint_{NL} dV \varepsilon_0 \sum_{ijk}
\chi^{(2)}_{ijk} E_{1i}^{*}  \left(E^{*}_{2j} E_{3k} + E^{*}_{2k}E_{3j}\right)}
{\sqrt{\iiint dV \varepsilon |\vec{E_1}|^2} \sqrt{\iiint dV \varepsilon |\vec{E_2}|^2} \sqrt{\iiint dV \varepsilon |\vec{E_3}|^2}}\right|^2
\end{align}

where the subscript $NL$ indicates that the integral in the numerator is taken over the region where there is nonlinear material present.  In this case, the overlap of the three modes involved (two in the case of SHG, with the pump mode appearing twice) must be even to have a non-zero interaction, and there must be coupling between the relevant polarizations via the $\chi^{(2)}$ nonlinearity (i.e. $\chi^{(2)}_{ijk} \neq 0$).

\begin{figure}
\includegraphics[width = 12cm]{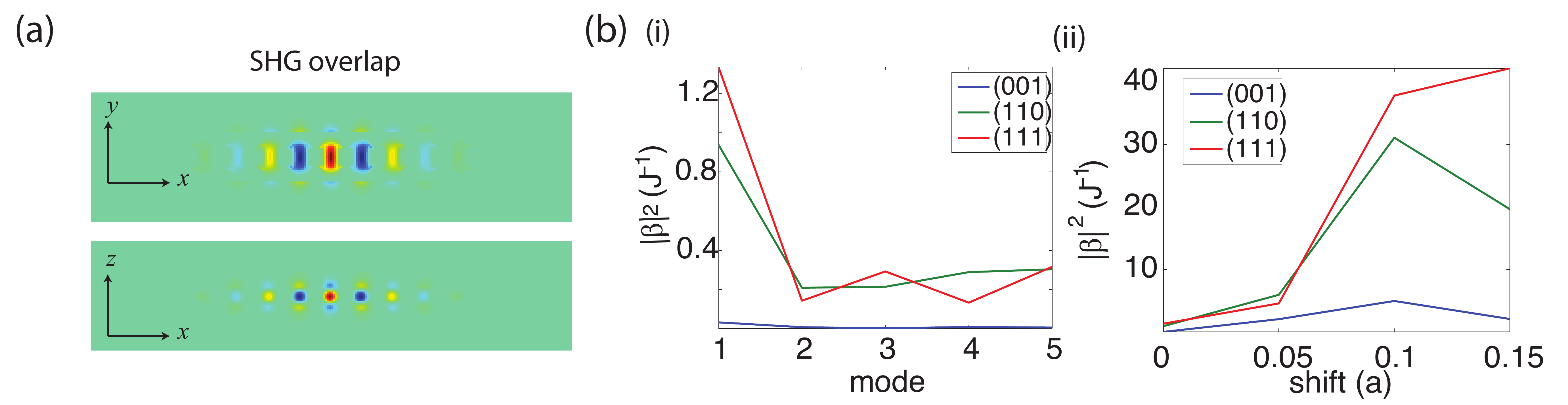}
\centering
{\caption{(a) The spatial profile of the nonlinear overlap between the $E_y$ field components of the modes shown in Fig. \ref{fig:SHG_mode1} (a) part (i) and Fig. \ref{fig:SHG_mode1} (b).  (b) (i) The nonlinear overlap for each of
the five modes shown in Fig. \ref{fig:SHG_mode1} (a) with the mode in Fig. \ref{fig:SHG_mode1} (b) for three different GaAs orientations. Part (ii) shows that if a cavity shift is introduced the nonlinear overlap can be significantly increased, but the Q factors of the modes will also be significantly decreased.}
\label{fig:SHG_mode_nonlin}}
\end{figure}

The spatial overlap for the two modes shown in Fig. \ref{fig:SHG_mode1} (a) part (i) and Fig. \ref{fig:SHG_mode1} (b), $E_{1y}^2E_{2y}$, is shown in Fig. \ref{fig:SHG_mode_nonlin} (a).  The overlap calculated from Eq. \ref{eq:beta} of the TE$_{002}$ localized mode with each of the five TE$_{000}$ localized modes is shown in Fig. \ref{fig:SHG_mode_nonlin} (b) part (i) for three different substrate orientations.  For each orientation, a different effective $\chi^{(2)}$ can be calculated, as shown in Appendix \ref{app:100_vs_111}, to be used in Eq. \ref{eq:beta}. As expected, the overlap is the lowest for the (001) orientation, where TE-TE mode coupling is not allowed. The overlap is also the highest with the lowest $V$ mode. However, the overlap is still relatively low.  We can compare the overlap to the maximum allowed overlap for a cavity with this mode volume, $\beta_{max} = \frac{\chi^{(2)}}{2n_1^2n_2\sqrt{\varepsilon_0V_2}} = 2.25 \times 10^3$.  As can be seen from Fig. \ref{fig:SHG_mode_nonlin} (a), the field overlap is approximately evenly divided between positive and negative field components, which cancel; this  explains the discrepancy between the simulated overlap and the maximum allowed overlap for these mode volumes. As shown in Fig. \ref{fig:SHG_mode_nonlin} (b) part (ii), we can improve the overlap by shifting the two holes near the cavity by 20 nm which improves the localization; however, this immediately decreases the Q of both the fundamental and SHG modes.

High conversion efficiency can also be achieved using two high Q modes.  In most previous demonstrations \cite{mccutcheon_experimental_2007,rivoire_second_2009, rivoire_sum-frequency_2010, Galli2010, buckley_second_2013, Ota2013:Nanocavity-based, Ota2013:Self-frequency,Diziain2013, Yamada2014, Buckley2014a, Buckley2014:Nonlinear}, these two modes were high Q input modes.   In this case, the output mode is a higher order leaky air band mode of the photonic crystal (above the light line).  Since there is a high density of such modes, a mode with the appropriate symmetry will always be found.  Therefore in this case the symmetry of the two input modes is not important, as the efficiency of the process will depend on the overlap of the two modes with a third, poorly defined mode (which will not necessarily improve with optimization of the overlap of the two modes \cite{Ota2013:Self-frequency}).

Another option is to have a cavity mode for one input and one output mode with the second input mode localized in a much larger cavity with approximately constant field \cite{Burgess2009:Difference-frequency, burgess_design_2009}, or as a focused Gaussian \cite{mccutcheon_broadband_2009}.  In this case, since the field in the third mode is approximately constant, the expression in Eq. \ref{eq:beta} can be simplified.  The exact form of the simplification depends on the symmetry of the $\chi^{(2)}$ tensor and the polarization of the constant field, but in the case of the (001) GaAs crystal orientation where the mode at $\omega_1$ is TE polarized and the mode at $\omega_2$ is TM polarized, this can be approximated \cite{Burgess2009:Difference-frequency,mccutcheon_broadband_2009,rivoire_multiply_2011}  as 

\begin{align}
\label{eq:gamma}
\gamma = \left|\frac{\varepsilon \iiint_{NL} \left(E_{1x}E_{2z}+E_{2z}E_{1y} \right) dV}{\sqrt{\iiint\varepsilon|\vec{E_1}|^2 dV}\sqrt{\iiint\varepsilon|\vec{E_2}|^2 dV}}\right|
\end{align}
where we have assumed a constant field with both $E_x$ and $E_y$ polarizations (further simplification can be made by choosing the precise polarization of the pump).  In order for $\gamma$ to be non-zero, the overlap of the two modes involved must have even symmetry in all directions.  There may also be additional requirements in choosing the appropriate mode. For example, InAs/GaAs quantum dots grown in the center of the GaAs membrane couple predominantly to TE-like modes.  By looking at the spatial symmetry of the products of the field profiles in Fig. \ref{fig:band_diag_modes2}, we can guess which combinations will likely have high $\gamma$ when confined as cavity modes.  For example, the TE$_{000}$ and TM$_{000}$ modes should have a very high nonlinear overlap, as shown in \cite{zhang_ultra-high-q_2009}. 

Using Eq. \ref{eq:gamma}, we can calculate a nonlinear figure of merit for each of these modes with the first TE$_{000}$ mode.  The TM$_{030}$ mode has the smallest overlap with the TE$_{000}$ mode, with the maximum figure of merit from Eq. \ref{eq:gamma} $\gamma = 0.0013$.  For the TM$_{020}$ mode, the overlap is much improved, with an overlap of 0.146. In the case of the TE$_{110}$ mode, as expected due to symmetry the overlap in Eq. \ref{eq:gamma} is close to zero.  

However, in the (110) or (111) GaAs crystal orientations (or for a different material system), due to the rotation of the photonic crystal axes relative to the electronic axes, the symmetry of the effective $\chi^{(2)}$ tensor changes \cite{Buckley2014a} such that TE-TE mode coupling is allowed (see Appendix \ref{app:100_vs_111}). In this case, the simplest form for the figure of merit $\gamma$ is
 \begin{align}
\label{eq:gamma111}
\gamma = \left|\frac{\varepsilon \iiint_{NL} \left(E_{1x}E_{2y}+E_{2x}E_{1y} \right) dV}{\sqrt{\iiint\varepsilon|\vec{E_1}|^2 dV}\sqrt{\iiint\varepsilon|\vec{E_2}|^2 dV}}\right|
\end{align}
where we have once again assumed interaction with a constant field with both $E_x$ and $E_y$ polarizations.   In these crystal orientations we expect to find that the first confined TE$_{000}$ mode has high overlap with the first confined TE$_{110}$ mode, and from Eq. \ref{eq:gamma111} we find an overlap of 0.52.  This is similar to overlaps achieved in TE/TM nano beams \cite{zhang_ultra-high-q_2009}.  The TE$_{002}$ mode has an overlap of 0.063, similar to the overlap achieved in crossbeam cavities \cite{Rivoire2011:Multiply2}, but with a much larger frequency separation and maintaining similar Q factors.  

Further improvement on these designs could use inverse design \cite{lu_nanophotonic_2013} or genetic algorithms \cite{Minkov2014} to optimize these resonances at widely spaced frequencies.  Even if the conversion efficiency per element is low, such doubly resonant microcavities could be linked together in a coupled cavity array to take advantage of both resonant cavity and slow light effects \cite{xu_propagation_2000}.

To achieve efficient frequency conversion, it is also important to engineer input and output coupling ports \cite{bi_high-efficiency_2012}. For nanobeam PCCs, this can be done via a side-coupled waveguide \cite{groblacher_highly_2013}.  Since the Q factor of the fundamental mode is so much higher than the Q of the second harmonic mode, the waveguide can be chosen to be critically coupled to the input mode without perturbing the second harmonic mode significantly.  Outcoupling of the second harmonic mode can be done via free space, or via a second waveguide with mirrors that prevent propagation at the fundamental wavelength but allow propagation at the second harmonic wavelength.    

\section{Linear characterization of structures}
\label{section:exp}

\begin{figure}
\includegraphics[width = 12.5cm]{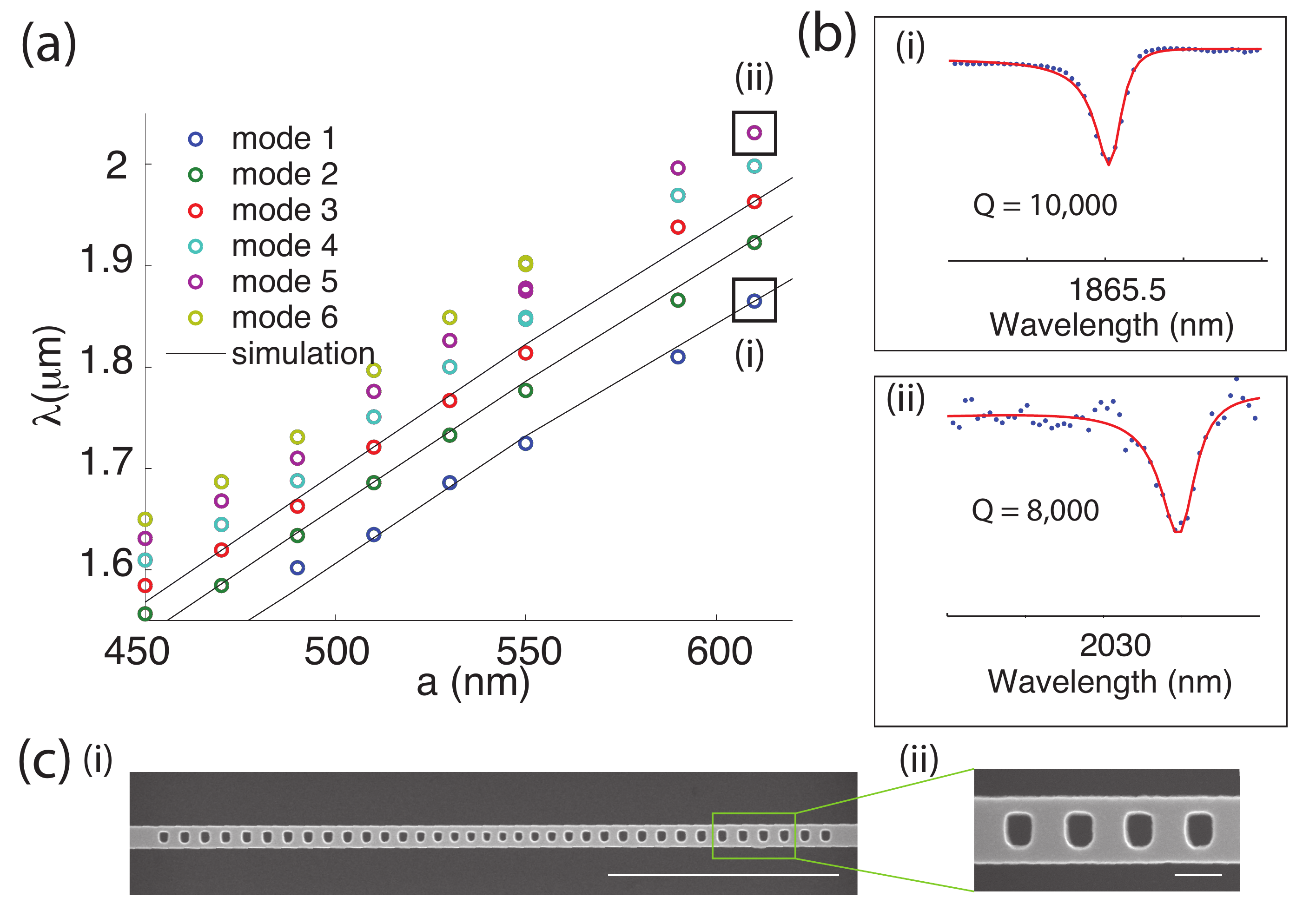}
\centering
{\caption{(a) Measured mode wavelength versus lattice constant for 8 different fabricated structures. 5-6 modes were measured for each structure, corresponding to the modes in Fig. 3.  (b) Representative spectra for two of the modes measured by fiber taper, with loaded Q factors of 10,000 and 8,000. (c) SEM images of a typical nanobeam with lattice constant 480 nm.  Scalebars are (i) 5 $\mu$m and (ii) 500 nm}
\label{fig:TE000}}
\end{figure}

We fabricate nanobeams in a 250 nm thick membrane (111)B oriented GaAs. The holes are patterned along nanobeams with lattice constants $a$ from 450 nm to 650 nm and beam widths from 1.07 to 1.5 $a$ as described previously \cite{buckley_second_2013}.  The rest of the design follows Section \ref{section:optimization}.  The highest frequency localized resonances of the TE$_{000}$ band of the nanobeam (which we will call the fundamental resonance) span from 1.45 $\mu$m to 1.87 $\mu$m with lattice constant variation, with the longest wavelength mode (the fifth confined mode of the band) measured at 2.03 $\mu$m.  Resonances up to 1.65 $\mu$m could be characterized via cross-polarized reflectivity  \cite{buckley_second_2013}. However, longer wavelength modes were characterized via fiber taper probing \cite{shambat_coupled_2010}, as the cross-polarized reflectivity method did not have high enough signal to noise.  In each case the cavity was probed with a broadband tungsten halogen white light source and detected on an extended InGaAs spectrometer.  The wavelengths of these five confined modes for fabricated parameters are shown versus lattice constant in Fig. \ref{fig:TE000} (a). Simulation results for the first three confined TE$_{000}$ modes are shown as the solid black lines in the figure. There is good agreement between simulation and experiment. The Q factors for the modes, as shown in Fig. \ref{fig:TE000} (b), are limited by coupling to the fiber taper, as indicated by the difference in Q factors measured via the cross polarized reflectivity method and the fiber taper method.  An example SEM of a nanobeam cavity is shown in Fig. \ref{fig:TE000} (c), with lattice constant 480 nm.  

\begin{figure}
\includegraphics[width = 10cm]{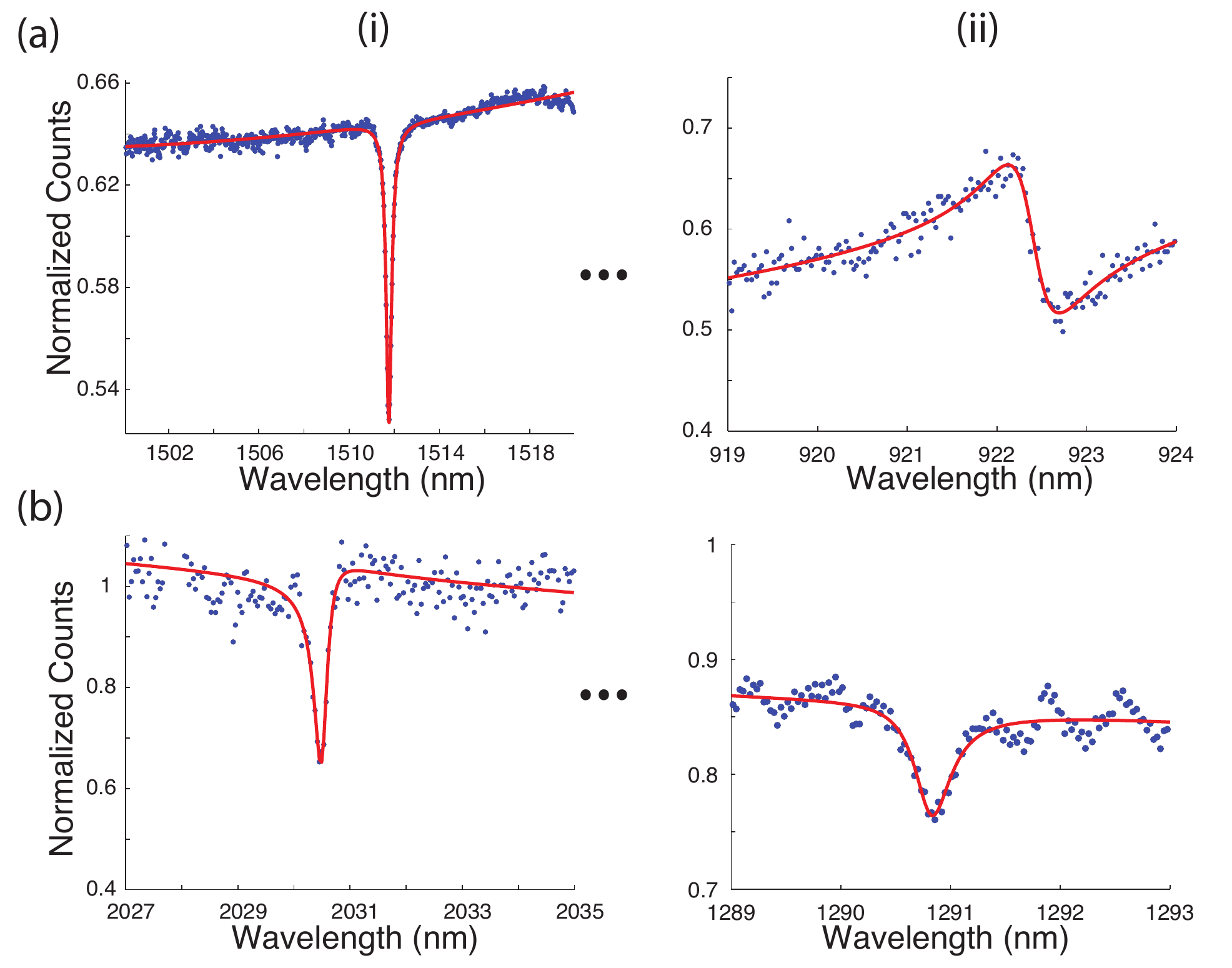}
\centering
{\caption{Demonstration of modes with large frequency separations.  (a) The fundamental TE$_{000}$ mode and a higher order mode at 920 nm, for a nanobeam with lattice constant 450 nm.  (b) The fifth confined TE$_000$ mode and first confined TE$_{110}$ mode of a nanobeam with lattice constant 610 nm.}
\label{fig:HOmodes}}
\end{figure}

We also characterized the higher order modes of these structures. We observed Q factors as high as 8,000 at 1300 nm for these higher order modes. However, in this case we could not observe the TE$_{000}$ mode as it was at a longer wavelength than we could detect. Since the membrane thickness is $t/a \approx 0.4$, we do not expect to see the TE$_{002}$ mode.   We also optimized designs with resonances at around telecommunications wavelengths and at 930 nm, the typical wavelength of InAs quantum dots, as shown in Fig. \ref{fig:HOmodes} (a).  The resonance at 1511 nm is the first confined TE$_{000}$ mode of the structure, which has the highest overlap $\gamma$ with the higher order mode, and was measured using a fiber taper.  The resonance at 930 nm was measured via cross polarized reflectivity, and is likely the first confined TE$_{110}$ mode of the structure.  This structure had a lattice constant of 450 nm.  Such structures could prove useful for frequency conversion between InAs QDs and the telecommunications network.  The largest mode separation we observed was 740 nm, for a nanobeam with lattice constant 610 nm (all other parameters as described in Section \ref{section:optimization}).  The measured resonances are shown in Fig. \ref{fig:HOmodes} (b), with the longest wavelength mode at 2.03 $\mu$m measured using a fiber taper, and two closely spaced higher order modes observed at 1293 nm and 1307 nm measured via cross polarized reflectivity (only the mode at 1307 nm is shown here). In this case, the longer wavelength mode is the fifth confined TE$_{000}$ mode, while the higher order modes are the first two confined (even and odd) TE$_{110}$ modes, as shown in Fig. \ref{fig:DFG_modes} (b), and also described in \cite{Buckley2014:Nonlinear}.

\section{Conclusions}
We demonstrated the design, fabrication and characterization of nanobeam cavities with multiple higher order modes.  Designs with two high Q modes with frequency separations of an octave were introduced, and we fabricated nanobeam cavities with frequency separations of up to 740 nm, corresponding to a factor of 1.57.  Such PCCs could have applications for frequency conversion of light from quantum emitters.

\section*{Acknowledgements}

Financial support was provided by the Air Force Office of Scientific Research, MURI Center for multi-functional light-€"matter interfaces based on atoms and solids, the National Science Foundation Grant 1406028, National Science Foundation Graduate Fellowships, and Stanford Graduate Fellowships. This work was performed in part at the Stanford Nanofabrication Facility of NNIN supported by the National Science Foundation under Grant No. ECS-9731293, and the National Science foundation Grant 1406028, and at the Stanford Nano Center. J.V. also thanks the Alexander von Humboldt Foundation for support.

\appendix

\section{Semiconductor Orientation}
\label{app:100_vs_111}

Following the convention used in \cite{Boyd2003}, we define the tensor, $d_{ijk}$ for a lossless medium away from resonance (where there is negligible dispersion of the susceptibility), 

\begin{align}
\label{eq:dijk}
d_{ijk} = \frac{1}{2}\chi^{(2)}_{ijk}
\end{align}

\noindent The nonlinear polarization can then be written as 

\begin{equation}
\label{eq:pi}
P_i(\omega_n+\omega_m) = \varepsilon_0\Sigma_{jk}\Sigma_{nm}2d_{ijk}E_j(\omega_n)E_k(\omega_m)
\end{equation}

For (100) oriented III-V semiconductors, the second order nonlinear
susceptibility tensor $d_{eff}$ is given by
\begin{equation}
d_{eff} =
\left( \begin{array}{cccccc}
0 & 0 & 0 & d_{41} & 0 & 0 \\
0 & 0 & 0 & 0 & d_{41} & 0 \\
0 & 0 & 0 & 0 & 0 & d_{41} \end{array} \right)
\end{equation}
where the generated polarization at the second harmonic can be calculated by
\begin{equation}
\left( \begin{array}{c}
P_x(2 \omega) \\
P_y(2 \omega) \\
P_z(2 \omega) \\ \end{array} \right) = 2 \varepsilon_0 d_{eff}\left( \begin{array} {c}
E_x(\omega)^2 \\
E_y(\omega)^2 \\
E_z(\omega)^2 \\
2E_y(\omega)\cdot E_z(\omega) \\
2E_x(\omega)\cdot E_z(\omega) \\
2E_y(\omega)\cdot E_x(\omega) \\ \end{array} \right)
\label{eq:d_100}
\end{equation}

For GaAs the $d_{eff}$ matrix found in the literature is defined such that
$x, y, z$ are the [100], [010] and [001] axes of the crystal structure
(while \{110\} are the cleavage planes).  This
means that for the case of (001) GaAs the $x$ and $y$ coordinates are in the
same plane as two of the major crystal axes, and can be chosen to be aligned
with the crystal axes.  Examining the mode overlap integral reveals that for
the process of second harmonic generation, a TE-like mode may only couple to
a TM-like mode if the wafer is normal to the [100], [010], or [001]
(equivalent) directions, as in standard (001) oriented wafers.

In the case of (111) GaAs, the plane of the wafer is no longer the same as
the plane of crystal axes.  The values of E-field can be either transformed
to this coordinate system or a new $d_{eff}$ matrix can be derived with
$x'$, $y'$ in the plane of the wafer.  Rotating from the (001) to the (111)
plane can be done by applying the following steps: (1) rotation about the
z-axis of 45 degrees (2) rotation through an angle of
$\arccos{\left(\frac{1}{\sqrt{3}}\right)}$ about the y-axis.

This gives the $d_{eff}$ matrix
\begin{equation}
\label{eq:d_111}
\left( \begin{array}{c}
P'_x(2\omega) \\
P'_y(2\omega) \\
P'_z(2\omega) \\ \end{array} \right) =
2 \varepsilon_0 d_{eff,111}\left( \begin{array} {c}
E_x(\omega)^{\prime2} \\
E_y(\omega)^{\prime2} \\
E_z(\omega)^{\prime2} \\
2E'_y(\omega)\cdot E'_z(\omega) \\
2E'_x(\omega)\cdot E'_z(\omega) \\
2E'_y(\omega)\cdot E'_x(\omega) \\ \end{array} \right)
\end{equation}
The calculated $d_{eff,111}$ is given by
\begin{equation}
d_{eff} = \left( \begin{array}{cccccc}
-\frac{1}{\sqrt{6}} & \frac{1}{\sqrt{6}} & 0 & 0 & -\frac{1}{\sqrt{3}} & 0 \\
0 & 0 & 0 & -\frac{1}{\sqrt{3}} & 0 & -\frac{1}{\sqrt{6}} \\
-\frac{1}{2\sqrt{3}} & -\frac{1}{2\sqrt{3}} & \frac{1}{\sqrt{3}}& 0 & 0 &
0 \end{array} \right)\cdot d_{41}
\end{equation}
We can see that this new $d_{eff}$ matrix leads to coupling between TE-like
polarizations, e.g. $P'_y = -\frac{1}{2\sqrt{3}} d_{41} E'_y E'_z
-\frac{1}{\sqrt{6}} d_{41} E'_y E'_x$. The tensor leads to the expected 120
degree symmetry, with the effective $x$ and $y$ axes along the
[11$\overline{2}$] and [$\overline{1}$10] directions.

Similarly for the (110) orientation, the calculated $d_{eff,110}$ is given by
\begin{equation}
d_{eff} = \left( \begin{array}{cccccc}
0 & 1/2 & -1/2 & 0 & 0 & 0 \\
0 & 0 & 0 & 0 & 0 & -1 \\
0 & 0 & 0 & 0 & -1 & 0 \end{array} \right)\cdot d_{41}
\end{equation}

with the effective $x$ and $y$ axes along the [00$\overline{1}$] and [1$\overline{1}$0] directions.

\end{document}